# Investigating Herculaneum papyri:

# An innovative 3D approach for the virtual unfolding of the rolls

Inna Bukreeva[1,2*] Michele Alessandrelli[3] Vincenzo Formoso[4] Graziano Ranocchia[3] Alessia Cedola[1]

[1] Consiglio Nazionale delle Ricerche, Istituto di Nanotecnologia, Rome Unit, I-00195 Rome, Italy
[2] P. N. *Lebedev* Physical *Institute*, Russian Academy of Sciences, Leninskii pr., 53 *Moscow, Russia*
[3] Consiglio Nazionale delle Ricerche, Istituto per il Lessico Intellettuale Europeo e Storia delle Idee, I-00161 Rome, Italy
[4] Università della Calabria, Dipartimento di Fisica, I-87036 Arcavacata Di Rende (Cosenza), Italy

* Corresponding author

**Abstract**

After the first recent attempts at virtually opening and reading Herculaneum papyri, a new enhanced method for virtual unfolding and peeling of ancient unopened papyrus rolls has been developed. This new algorithms-based and semi-automatic procedure allowed to investigate with high resolution the 3D internal structure of two ancient papyrus rolls formerly analysed by X-ray phase-contrast tomography. Through a new rendering procedure, selected regions inside the rolls could be identified and isolated and the different sheets could be peeled one by one virtually, all without damaging these very precious manuscripts. Subsequently, traces of possible text were identified on the single sheets. Finally, by applying an additional flattening procedure we restored for the first time several extensive textual portions of Greek, the largest ever detected so far in unopened Herculaneum papyrus rolls, with different degrees of resolution and contrast.

# INTRODUCTION

As is known to specialists, one of the most demanding and promising tasks of papyrology today is to open and read virtually the Greek and Latin texts contained in the Herculaneum papyrus rolls, a collection of about 1840 catalogued papyri mostly stored in the National Library of Naples, which were carbonised and buried by the eruption of Mount Vesuvius in 79 AD until their re-discovery in the XVIII century. This unique library, the only one to have directly survived from antiquity, includes some hundreds of still unopened bookrolls, which very probably hand down inedited treasures of ancient philosophy and classical literature. After mechanical, chemical and electro-magnetical systems were employed to this end in the past three centuries – not always successfully and with either partial destruction or severe fragmentation – X-ray phase-contrast tomography (XPCT) has recently proven to be a suitable way for unwrapping and reading these very precious manuscripts in a non-disruptive way. In December 2013 this technique was applied for the first time to a Herculaneum papyrus roll and a Herculaneum papyrus fragment both owned by the Institute of France (Paris). This feasibility test experiment showed the potential benefit of applying XPCT to ancient papyri, but neither virtual unrolling nor extensive text reading was achieved [1]. In September 2015 another experiment was performed by our team to two Herculaneum papyri owned by Naples' National Library, and a set of numerical algorithms for 'virtual unrolling' was developed in-house leading to the revelation of some portions of Greek text inside both papyrus rolls, with unprecedented spatial resolution and contrast [2].

The present paper investigates in detail the complex 3D spatial behaviour of the sheets inside *PHerc*. 375 and *PHerc*. 495, two Herculaneum papyrus rolls owned by the National Library of Naples and formerly analysed by us through XPCT at the *European Synchrotron Radiation Facility* of Grenoble [2]. Through an *ad hoc* newly developed 3D rendering procedure, selected regions inside the rolls can now be identified and virtually isolated. Subsequently, the different packs of sheets are peeled one by one virtually reproducing, as it were, the historical method of the 'scorzatura'. Preliminary flattening of a volume-thick pack of papyrus which comprises a number of papyrus sheets significantly simplifies the segmentation of an isolated sheet or a stack of a few papyrus sheets. This helps to find traces of potential writing on the single sheets. As soon as a probable text is found, an additional flattening procedure contributes to restore the text. By this way, several textual portions of up to fourteen lines were identified for the first time, the largest ever detected so far in unopened papyrus rolls, with different degrees of resolution and contrast.

# RESULTS

**Herculaneum papyri after 79 AD eruption of Mount Vesuvius.**

During the eruption of Mount Vesuvius, papyri were not only carbonized, but also crushed, compressed and partially melted. Under the influence of high temperature and pressure, the regular cylindrical shaped papyrus rolls with spiral-wrapped sheets got a highly irregular configuration individual for each papyrus.

*PHerc*. 495, as imaged by X-ray Phase contrast Tomography in Fig.1a, resembles the cylinder with a depression in the central part. Sheets in the inner part of the papyrus are loosely wrapped around the centre sometimes in quite a regular way, as can be seen from the orthogonal cross section of the roll (see Fig.1b). However, in the longitudinal projection, sheets have a very complicated spatial orientation with sophisticated curved pleat geometry (see Fig.1c), which makes it difficult to discriminate and to separate the sheet's surface.

*PHerc*. 495 (see Figs.2) is noticeably compressed and corrugated with tightly folded and twisted sheets. By appearance, they resemble a thick stack of paper compressed and twisted with force (see Fig.2(c,d)). In some parts the papyrus is compressed so tightly that sheets, flattened by pressure, take a fairly regular form and become almost parallel to each other.

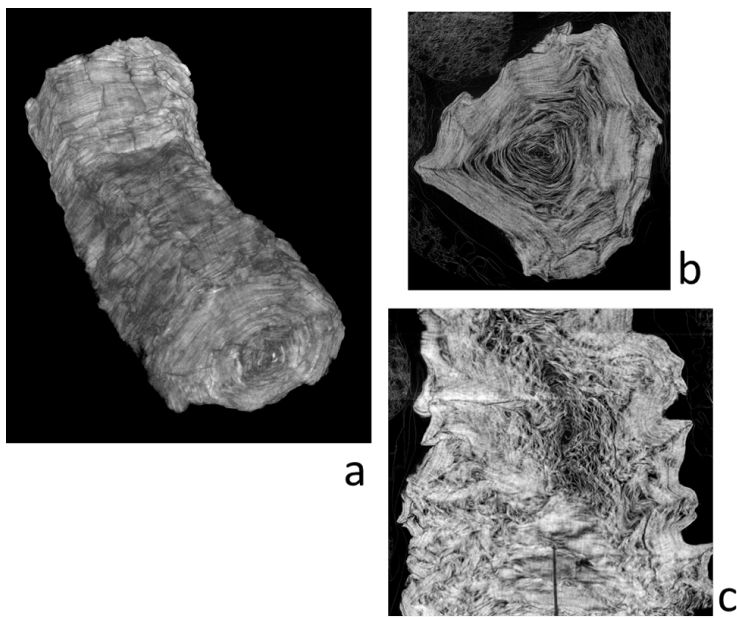

Fig.1 *PHerc*. 495 a) measured part; b) orthogonal cross section; c) longitudinal cross section.

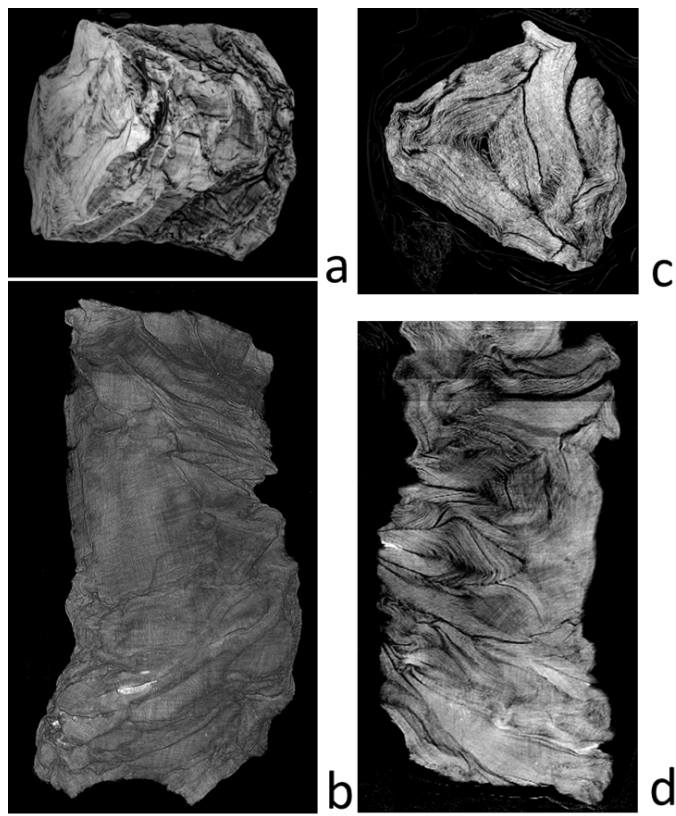

Fig.2 *PHerc*. 495 a) upper part; b) lower part; c) orthogonal cross section; d) longitudinal cross section.

Despite the significant difference in the internal structure of both rolls, generally, three main zones in each of them can be distinguished according to after-effect due to high temperature and pressure:

1. The main thermal shock which occurred on the exterior part of the roll resulted in a dense crust. In this portion, sheets are damaged to the utmost degree. In some places they are merged to form a quasi-monolith. The scrolls in addition are squashed from the upper and lower sides and their internal structure in these parts is similar to that of the other external areas. However, in most cases the roll has sufficiently ample areas with a rather regular shape maintaining a layered structure.

2. The interior part of the roll was less affected by the extreme external conditions than the rest of it. Sheets are more loosely rolled up around the centre. However, as a rule, sheets have a complex 3D spatial behaviour.

3. In the intermediate part, the sheets are mainly well preserved and in some areas they can take a regular form due to the tight compression.

Unwrapping of Herculaneum papyri with tightly compressed and tangled sheets requires approaches different from the methods used for well separated and loosely rolled up sheets of scrolls [3-5]. In this article we discuss a method applicable, in particular, to *PHerc*. 495, which has tightly packed sheets and to that part of the *PHerc*. 375, which was strongly compressed during the eruption. In particular, this approach is useful for the virtual unfolding of the intermediate and the exterior part of the rolls where, in particular, such important information as the author's name and the title of the work were usually located.

**Virtual unfolding**

The first step of the 3D data processing includes visualisation and analysis of the spatial behaviour of the sheets. Following the 3D rendering procedure, the region of interest inside the scroll can be identified. Figure 3a shows the 3D intricate longitudinal spatial configuration of the sheets inside the roll, while Fig.3b demonstrates that large areas of the sheets originally had an approximately regular or partially flattened shape due to the external pressure occurred during the eruption.

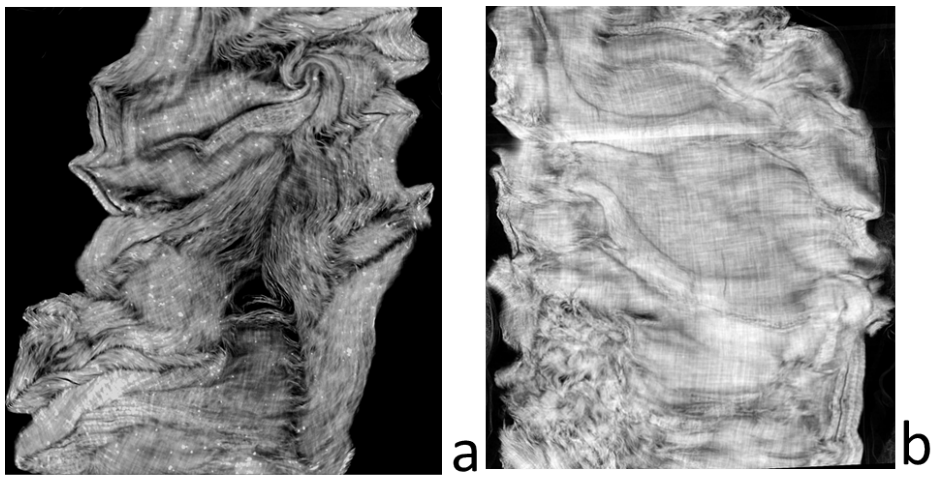

Fig.3 The spatial configuration of the sheets inside *PHerc*. 495 (upper part corresponding to Fig.2a). 3D rendering.

In order to prevent data loss or misinterpretation due to the complicated 3D spatial behaviour of the sheets we started exploring areas having a sheets configuration close to the developable surface, i.e. surface such as generalized cylinder, conical surface, trivially plane surface or/and combinations of these shapes that can be flattened onto plane without distortions. Subsequently, the method can be adapted to a more complex geometry.

After we have selected the region of interest, the area chosen for the study can be virtually separated and isolated. Figure 4 shows the virtually selected portion of *PHerc.* 375 before and after the virtual removal of the outer pack of sheets. The same procedure can be applied to separating and isolating any selected pack of sheets inside the roll.

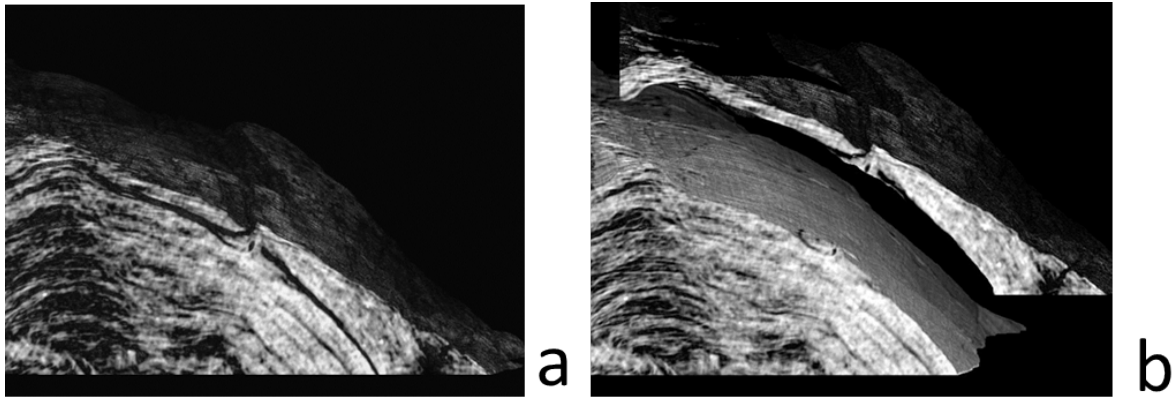

Fig.4 *PHerc.* 375. Virtual analogue of the mechanical removing of the outermost layers ('barks') from the external surface of the roll.

Figure 5a shows almost parallel compressed sheets in the external part of papyrus roll *PHerc.* 495 (top and middle part of the figure) and *PHerc.* 375 (down part of the figure). Figure 5b demonstrates the innermost part of *PHerc.* 495, around which the sheets were rolled up. Figure 5c delineates quasi-parallel sheets geometry in the intermediate part of *PHerc.* 495.

As far as the pack consists of tightly pressed quasi parallel sheets (see Fig.5), the unwrapping of the scroll virtually reproduces the same procedure as our historical ancestors mechanically performed in the XVIIIth and the XIXth century by a disruptive way ('scorzatura'): removing the outer barks ('scorze') of the roll and peeling the layers following the surface of sheets one by one. This procedure helps to find traces of potential handwriting within the pack of sheets. As soon as a probable text is found, the additional flattening procedure can help to restore the hidden text from the pack's volume.

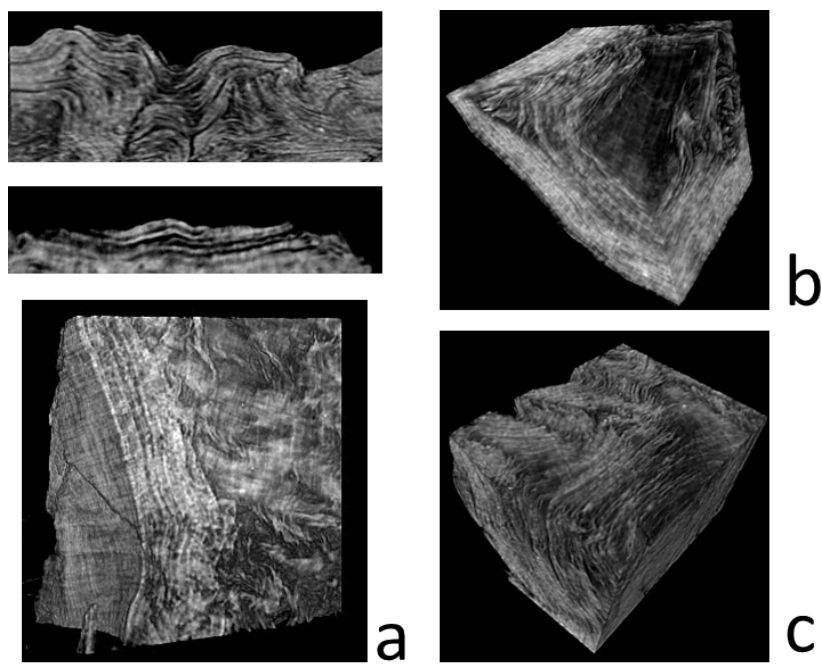

Fig.5 Internal structure of papyrus rolls *PHerc*. 495 and *PHerc*. 375: (a) top, middle: a cross section of the exterior part of the roll *PHerc*. 495 and (a) down: 3D view of the exterior part of *PHerc*. 375 with well pronounced periodical structure; (b) 3D view of *PHerc*. 495 shows the innermost part of the roll, around which the sheets were rolled up; (c) cross section of *PHerc*. 495 with quasi-parallel sheets in the intermediate part.

Figure 6 demonstrates subsequent steps of the virtual unfolding procedure of *PHerc*. 375. Figure 6a shows the spatial configuration of the sheets inside the chosen portion of the roll. In this figure, at least three different layers with cracks and pleats are clearly visible. The text traces detected with the virtual surface peeling procedure are shown in Fig.6b. Figure 6c shows the results of the additional flattening procedure.

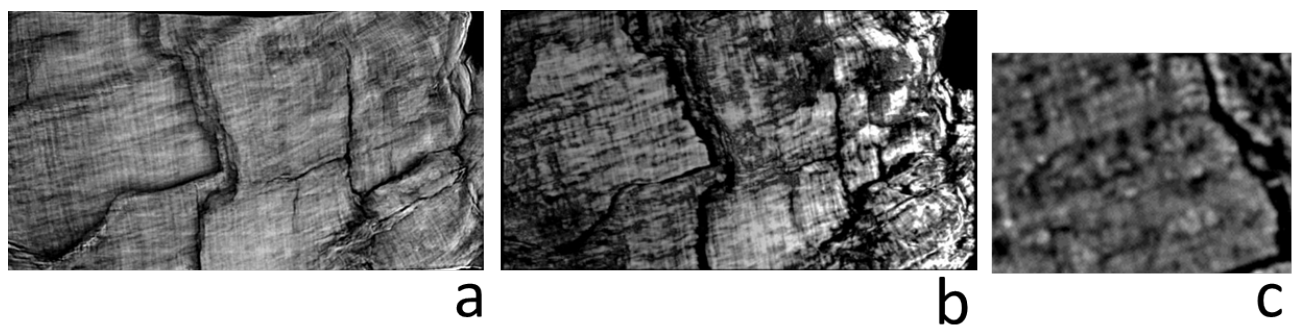

Fig.6 Internal structure of papyrus roll *PHerc*. 375: (a) at least three separate thick layers are clearly visible; (b) text traces are visible on the left layer; (c) restored content of the sheet.

As is shown in Fig.7 accurate flattening procedure is essential for the text detection. Figure 7a demonstrates strong distortions in the flattened image, which evidence the necessity of additional flattening procedure. Opportunely the criss-crossed fibres of the papyrus tissue represent the natural greed lines (see Figs.7b,c), which help to control the accuracy of the unfolding procedure . Figure 7d shows that accurate flattening procedure helps to reveal traces compatible with writing.

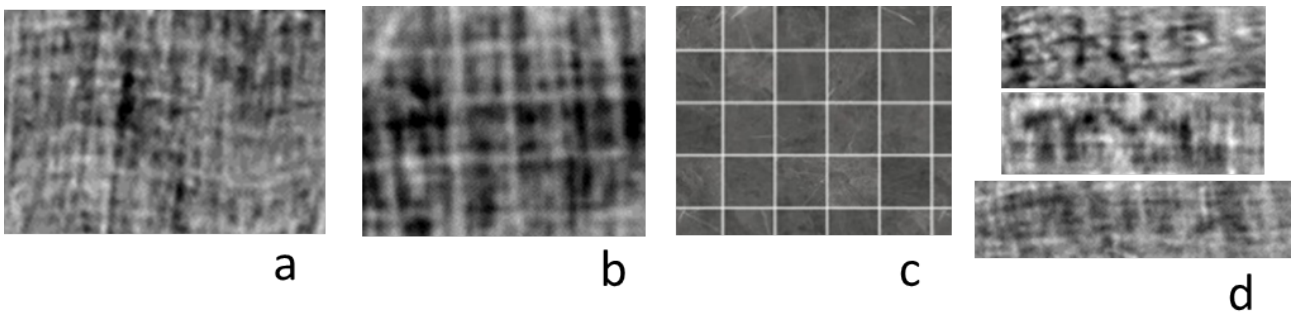

Fig.7 Results of the flattening procedure: (a) strong distortions in the flattened image; (b) the criss-crossed fibres of the papyrus; (c) greed lines; (d) additional flattening of the sheet helps to reveal traces of writing.

**Problems related to text detection**

Although the apparent simplicity of the method, many problems appear related to the complicated geometry of the sheets, the internal structure of the papyrus tissue and the numerous artefacts encountered in the investigation of the rolls. Irregular intrinsic structure of the sheets, pleats, holes, tears, external contamination (see Fig.8) tend to 'imitate' the text and these problems should be seriously considered when trying to identify possible text. Another possible source of error are extrinsic artefacts introduced during phase-contrast tomographic reconstruction and/or flattening procedure because in this case only approximate solutions are possible and algorithms can only aim to minimize geometric distortions, yet cannot prevent distortions altogether.

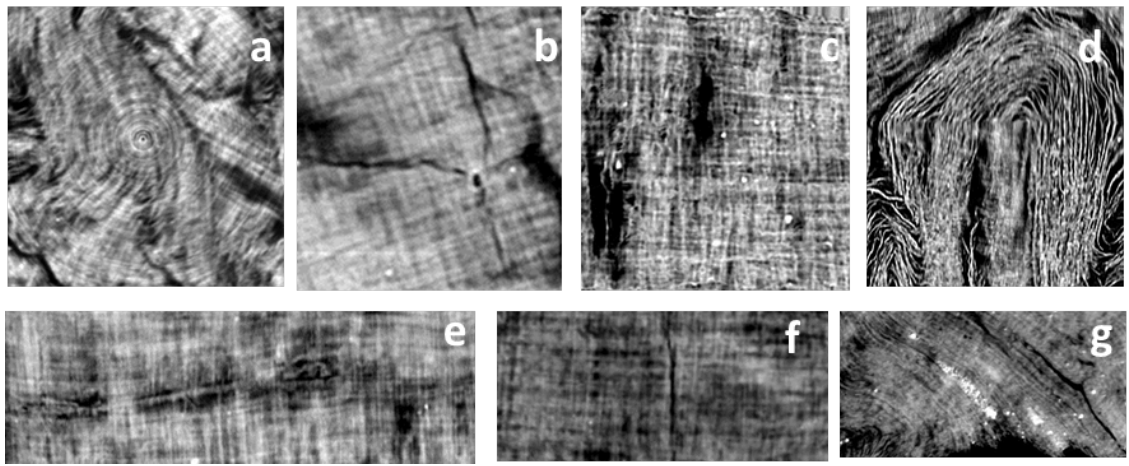

Fig.8 Artefacts and internal structure of papyrus rolls *PHerc*. 375 and *PHerc*. 495: (a) circular artefact in tomographic reconstruction, (b) hole and cracks arising from possible needles introduced into the rolls in modern times, (c) ruptures in a both papyri sheet due to extremal external condition, (d) complex bends and folds of the sheets; (e) structure of the sheet with horizontal fibres in *PHerc*. 495, (f) cracks on the sheet surface; (g) external contamination of the papyri, especially in *PHerc*. 495, by small stones and sand.

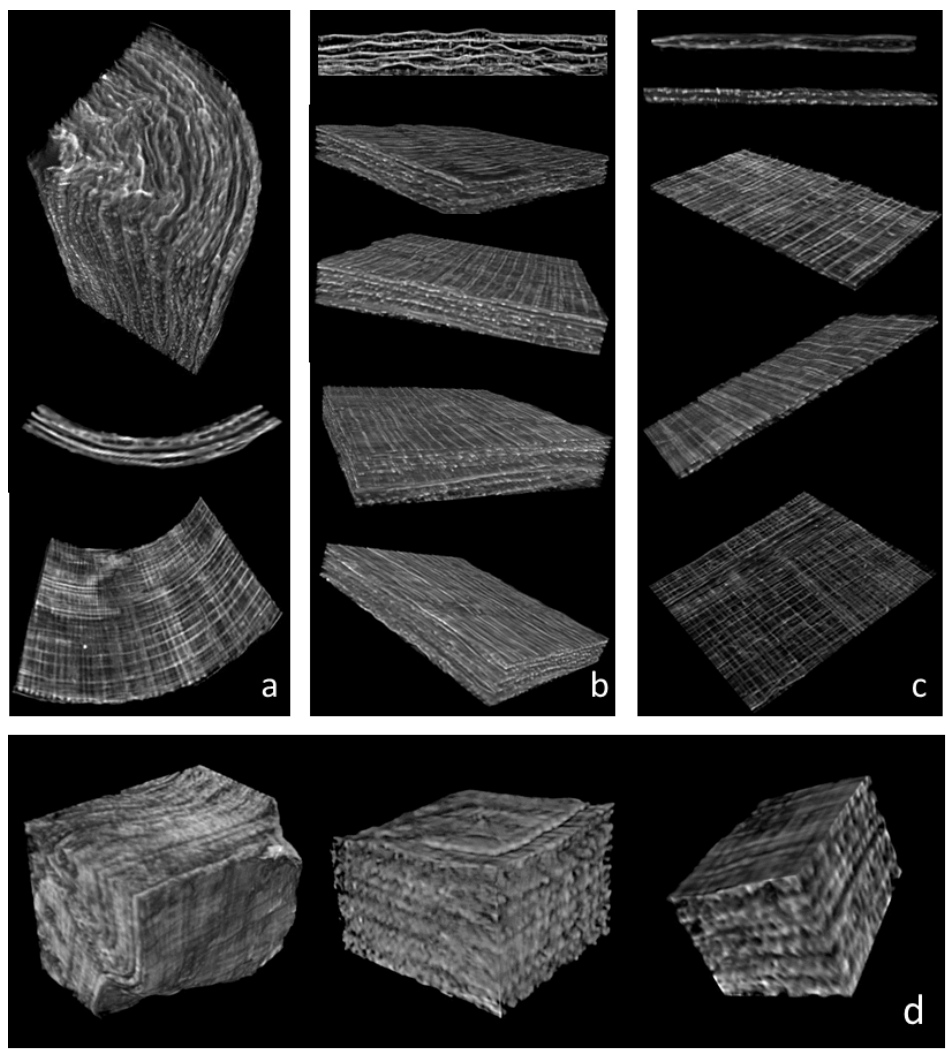

Fig.9 (a) top: the quadrant of the tomographically reconstructed 3D image of the papyrus-roll phantom, middle and down: 3D image of the stack composed with several folded sheets of papyrus; (b) 3D images of unfolded volume-thick slabs of the papyrus phantom; (c) 3D images of a thin layer of the papyrus segmented after the volume flattening procedure, (d) left: 3D view of a selected section of *PHerc. 495*, middle: selected and flattened part of the roll, right: the portion of the papyrus after removal of the external part.

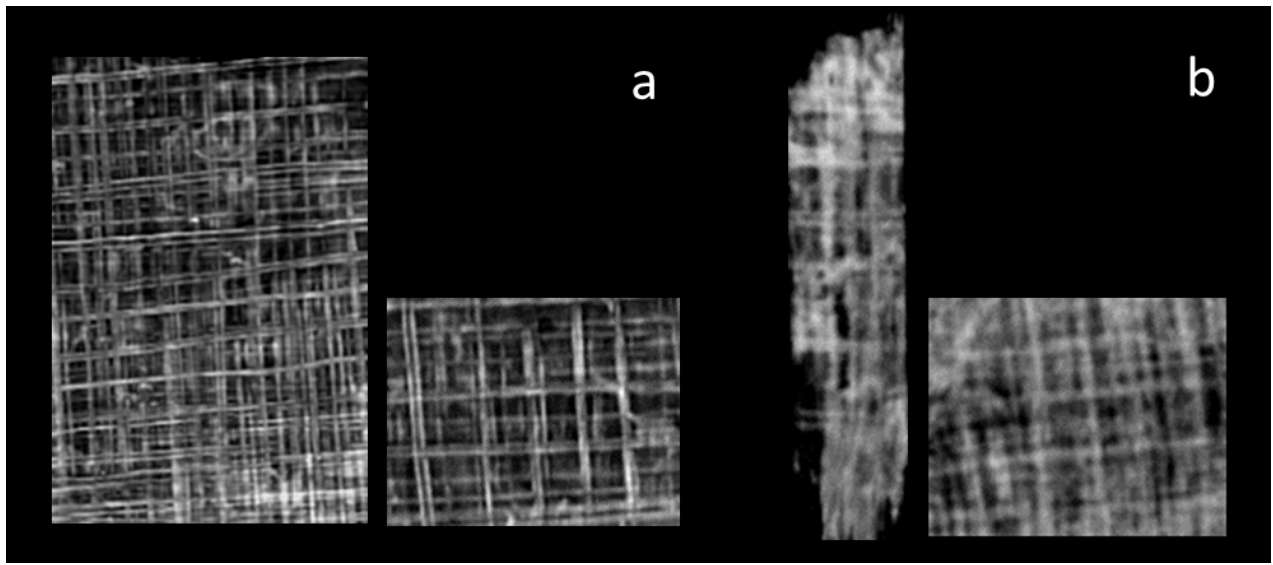

Fig.10 (a) The criss-crossed fibre structure of the sheets of the phantom papyrus; (b) left: the fibre texture of the sheet segmented from the innermost part of the original papyrus roll *PHerc*. 495, (b) right: the sheet of *PHerc*. 495 from the external part of *PHerc*. 495.

The phantom was fabricated and carbonized to reproduce ancient papyrus rolls in geometry, material and physical condition. Sequence of numbers and Greek letters were written on the paper sheets using home-made ink before rolling up. The procedure of the carbonized phantom fabrication, composition of the ink, description of the experimental conditions, as well as details of data processing and results of virtual unfolding and revealing of writing can be found in the Supplementary Information of the article [2]. The phantom was fabricated in order to facilitate the development and test of the virtual unfolding algorithms as well as for investigating the scroll's internal structure in a reliable way.

The quadrant of the tomographically reconstructed 3D image of the phantom is shown on the top of Fig. 9a. Below the tomographic images of the phantom, the stack composed of several folded sheets of papyrus is displayed. Figure 9b illustrates the result of the unfolding procedure of a volume-thick slab of the papyrus which comprises a number of papyrus sheets. The volume flattening procedure significantly simplifies the segmentation of an isolated papyrus sheet or a stack of a few sheets. Figure 9c shows 3D images of a thin papyrus layer segmented after the volume flattening procedure. 3D view of the portion of *PHerc*. 495 is shown in the left part of Fig.9d. A selected and flattened part of the papyrus is presented in the middle of Fig.9d. The papyrus portion after removal of the external part is demonstrated in the right part of Fig9d.

Figure 10 clearly shows the fibrous structure of the material with parallel fibers for each layer and right angle of the fibers direction for alternating layers. Fibers are well visible and can be used as reliable guidelines for the identification of the individual sheet. Moreover, they help to control the process of unfolding and flattening of the papyrus sheets.

Figure 10a shows the criss-crossed fibers structure of the layered sheets of the papyrus phantom. Figure 10b highlights the fibers texture of the original papyrus *PHerc*. 495 with periodicity of about 1-1.5 millimetres.

The papyrus sheet was basically made from stripes cut from stems of *Cyperus Papyrus* and, after suitable preparation, had a smooth surface which was significantly deformed and degraded under extreme external conditions. Moreover during carbonization of papyri due to the drastic volcano eruption, chemical transformation of the plant's elements and/or the external contaminants cannot be excluded. Both the plant material and paper sheets have a complex cellular structure [8,9], and therefore dust and other external contaminants can penetrate deeply into the stratum. In this situation, insufficient resolution of the detector as well as inaccuracy in flattening and 3D rendering procedures can bring misleading information about structure, contamination, ink etc. in the same pixel and can finally 'imitate' the text.

Therefore, in order to achieve a reliable differentiation of the text from the structure it is necessary to further deepen the study of the structure of carbonized papyrus and the composition of the ink. Moreover, the optimization of the experimental parameters including improved detector resolution, is required. The results will be the subject of a forthcoming publication.

**Textual discoveries**

The optimization of the virtual unfolding procedure, the differentiation of potential writing from the papyrus structure, allowed us to detect inside each papyrus roll the most extensive textual portions ever read so far with different degrees of contrast and resolution. The texts have always been found on the *recto*. Although in various cases the decipherment is still difficult to perform, the distribution of the lines appears to be evident. This is the case, for instance, of Figs.11 and 12, both taken from *PHerc*. 375. In Fig.11, three perfectly parallel text lines are visible. Several letters are legible. In Fig.12, up to fourteen lines are detectable and, albeit the text is not sufficiently clear, several letters and words might be conjectured. The distribution of the lines appears to be here more regular in the upper than the lower portion of the picture. This is due to the severe convolution and distortion of the papyrus layers inside the unopened roll prior to the virtual unfolding. The same applies to Fig.13, from *PHerc*. 495, where we isolated a text of up to ten lines. In this case too, the lines do not appear perfectly parallel. In this condition, every textual reconstruction would be haphazard. This notwithstanding, single letters and syllables are recognizable.

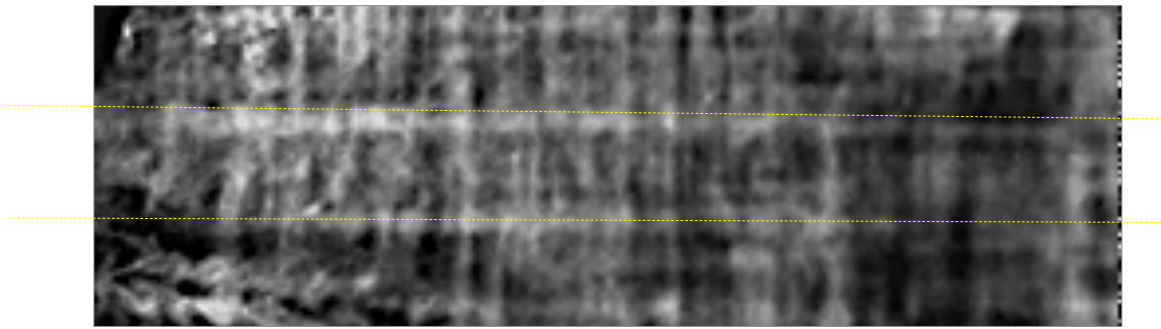

Fig.11 Three parallel text lines from *PHerc*. 375.

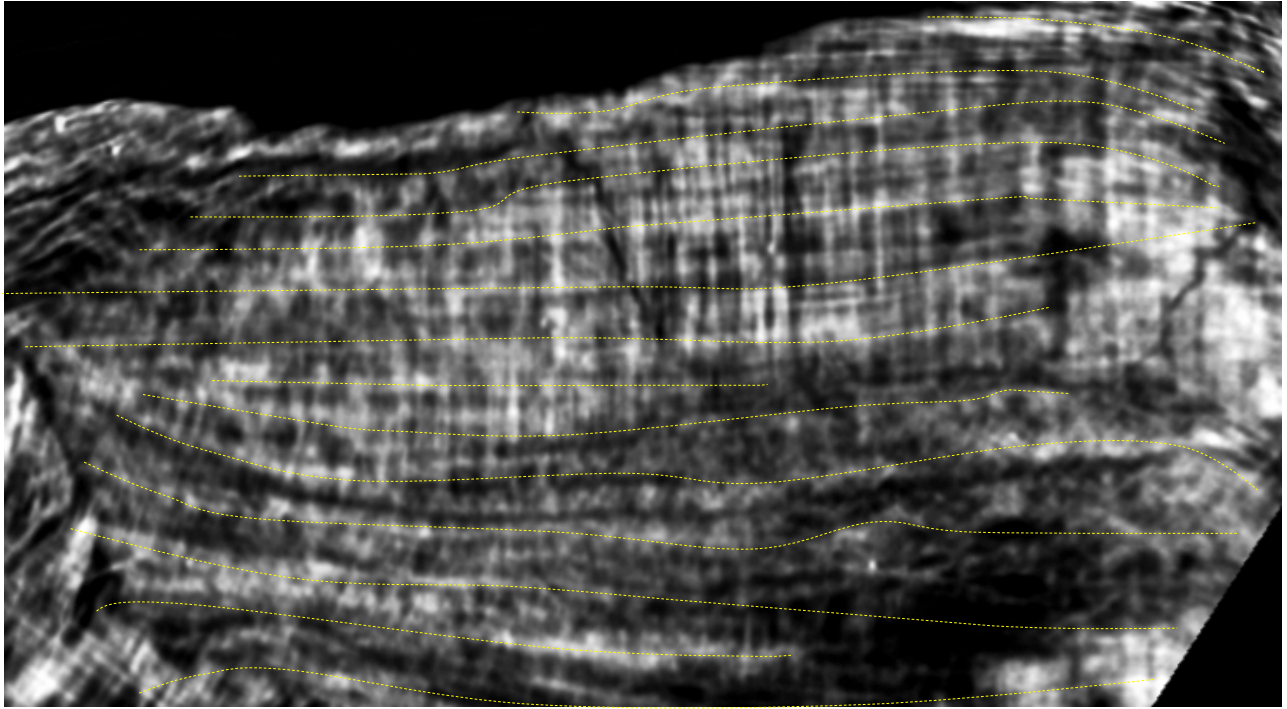

Fig.12 Several text lines from *PHerc.* 375.

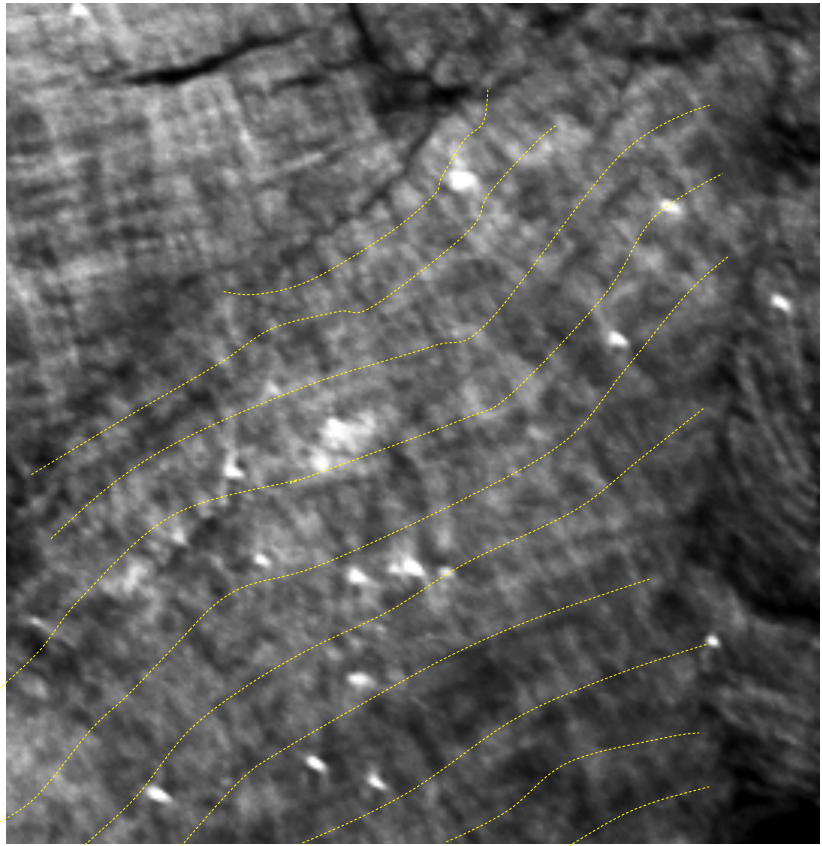

Fig.13 Several text lines from *PHerc.* 495.

In Fig.14, taken from *PHerc.* 495, seven quite regular lines are detectable and the text seems to be clearer. In particular, at line 2, a compound word in ἀπω[ like a voice of the verb ἀπωθέω, 'push back', 'drive away', or the noun ἀπώλεια, 'destruction', can be advanced. At line 6, either a word ending in ]τι followed by another beginning with ποι- or a voice of the verb ἀντιποιέομαι, 'to lay claim to', are equally possible. In Philodemus, in particular, this verb is especially attested in *On Rhetoric* (194, 23; 344, 11 Sudhaus I; 210, 27 Sudhaus II and also 205, 6 Sudhaus I) [10]. In Fig.15, also coming from *PHerc.* 495, the text, despite its lesser extension, is significantly more legible than in the previous cases. Contrast and resolution are here at their currently utmost degree. There appear to be remnants of three lines. At line 1, we possibly have ἀ]κọήν, 'hearing'..

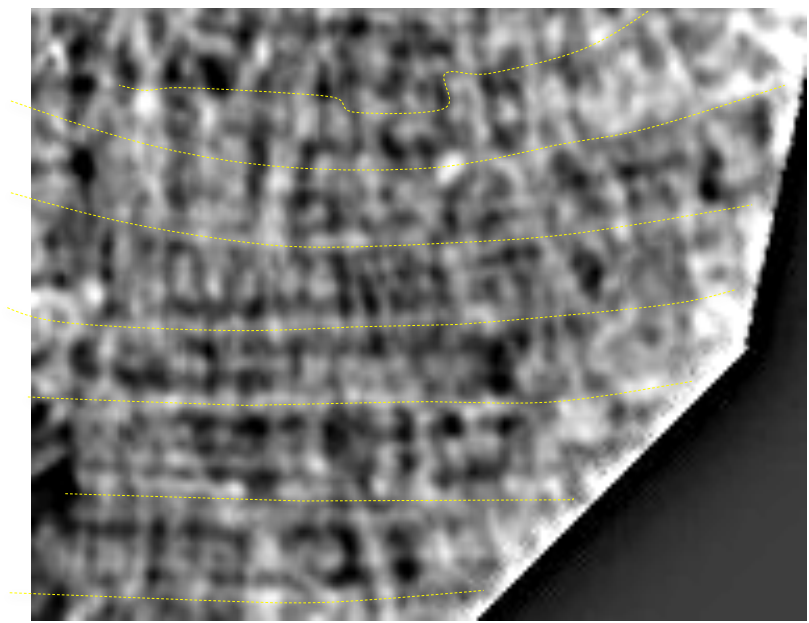

Fig.14 Seven text lines from *PHerc.* 495: letters and words recognizable.

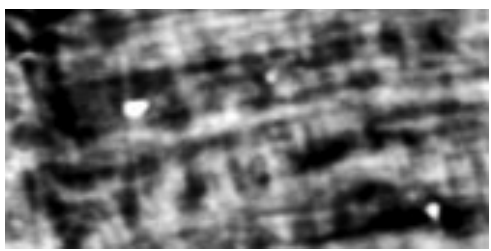

Fig.15 Three text lines from *PHerc.* 495: words recognizable.

The process of detection and decipherment of the text contained in both papyrus rolls continues on today producing progresses, which were initially unexpected. The optimisation and the automatisation of the virtual unfolding procedure open up new inedited perspectives in this direction, which are likely to produce a future breakthrough in our knowledge and understanding of ancient philosophy and classical literature.

## METHODS

Experiment were carried out at ID17 of the European Synchrotron Radiation Facility (ESRF) in Grenoble, using free-space propagation set-up. The image detector is Fast Read out Low Noise (FReLoN) 3k charge-coupled device camera (CCD) connected with an X-ray optics determining an effective pixel size of 47x47 μm. A double-silicon (111) crystal system was used to monochromatize the incident X-ray beam. Sample was located at about 10 meters from CCD camera and they were rotated around the vertical axis parallel to the longitudinal axis of the rolls. The X-ray phase contrast tomography (XPCT) experiments on the Herculaneum Papyri were performed at X-ray beam energy of 73 keV, the experiment on the phantom were performed at 54 keV and 80 keV. The spatial resolution is limited by pixel size.

## REFERENCES

[1] V. Mocella, E. Brun, C. Ferrero, D. Delattre, Revealing letters in rolled Herculaneum papyri by X-ray phase-contrast imaging. *Nat. Commun.* **6,** 1–5, doi: 10.1038/ncomms6895 (2015).

[2] I. Bukreeva, A. Mittone, A. Bravin, G. Festa, M. Alessandrelli, P. Coan, , V. Formoso, R. G. Agostino, M. Giocondo, F. Ciuchi, M. Fratini, L. Massimi, A. Lamarra, C. Andreani, R. Bartolino, G. Gigli, G. Ranocchia, A. Cedola, Virtual unrolling and deciphering of Herculaneum papyri by X-ray phase-contrast tomography, *Scientific Reports* **6**, Article number: 27227 (2016) doi:10.1038/srep27227

[3] D.Mills, O. Samko, P.L. Rosin, K. Thomas, T. Wess, G.R. Davis. Apocalypto: revealing the unreadable. Proc. SPIE 8506, Developments in X-Ray Tomography VIII, 85060 A (2012).

[4] O. Samko, Y.K. Lai, D. Marshall, P. Rosin, Virtual unrolling and information recovery from scanned scrolled historical documents. Pattern Recognit. 47(1), 248–259 (2014)

[5] G.H. Barfod, J.M. Larsen, A. Lichtenberger, R. Raja. Revealing text in a complexly rolled silver scroll from Jerash with computed tomography and advanced imaging software. Sci. Rep., 5:17765 (2015)

[6] W. B. Seales, C. S. Parker, M. Segal, E. Tov, P. Shor, Y. Porath. From damage to discovery via virtual unwrapping: reading the scroll from En-Gedi. Sci. Adv. 2:e1601247 (2016)

[7] D. Baum, N. Lindow, H.-C. Hege, V. Lepper, T. Siopi, F. Kutz, K. Mahlow, H.-E. Mahnke, Revealing hidden text in rolled and folded papyri, Appl. Phys. A 123:171 (2017)

[8] A. Waller, The Reconstruction of Papyrus Manufacture: A Preliminary Investigation, Studies in Conservation, Vol. 34, No. 1 (Feb., 1989), pp. 1-8 Published by: Taylor & Francis, Ltd. on behalf of the International Institute for Conservation of Historic and Artistic Works Stable URL: http://www.jstor.org/stable/1506154

[9] E. Franceschi, G. Luciano, F. Carosi, L. Cornara, C. Montanari, Thermal and microscope analysis as a tool in the characterisation of ancient papyri, Thermochimica Acta 418 (2004) 39–45

[10] S. Sudhaus, (ed.), *Philodemi volumina rhetorica*, Lipsiae, Teubner, I, 1892; II, 1895; Suppl., 1895.

## ACKNOWLEDGEMENTS

Thanks are due to the National Library of Naples 'Vittorio Emanuele III' for the precious advice on historical and conservation matters. The European project VOXEL "Volumetric Medical X-Ray Imaging at extremely low dose" (HORIZON 2020-Fet Open; Project reference: 665207) is acknowledged for financial support.